\documentclass[twocolumn,showpacs,preprintnumbers,amsmath,amssymb,aps]{revtex4-1}

\usepackage{graphicx}% Include figure files
\usepackage{dcolumn}% Align table columns on decimal point
\usepackage{bm}% bold math

\usepackage[T1]{fontenc} %inne kodowanie (to wlaczamy ja usuniemy [MeX]{polski}

\begin{document}

\preprint{Submitted to: ACTA PHYSICA POLONICA A}

\title{Magnetic orderings and phase separations in the zero-bandwidth limit \\
of the extended Hubbard model with intersite magnetic interactions}% Force line breaks with \\

\author{Waldemar K\l{}obus}
\author{Konrad Kapcia}%
    \email{kakonrad@amu.edu.pl}
\author{Stanis\l{}aw Robaszkiewicz}%
\affiliation{Electron States of Solids Division, Faculty of Physics, Adam Mickiewicz University,
ul. Umultowska 85, 61-614 Pozna\'n, POLAND}%

\date{July 5, 2010}% It is always \today, today, but any date may be explicitly specified

\begin{abstract}
A simple effective model for a description of magnetically  ordered insulators is analysed. The tight binding Hamiltonian consists of the effective on-site interaction ($U$) and intersite magnetic exchange interactions ($J^z$, $J^{xy}$) between nearest-neighbours.
The phase diagrams  of this model  have been determined within the variational approach, which treats the on-site interaction term exactly and the intersite interactions within the mean-field approximation.
We show that, depending on the values of interaction parameters and the electron concentration, the system can exhibit not only homogeneous phases: (anti-)ferromagnetic (F$_\alpha$) and nonordered (NO), but also phase separated states (PS$_\alpha$: \mbox{F$_\alpha$--NO}).
\end{abstract}

\maketitle

\section{Introduction}

The extended Hubbard model with anisotropic spin exchange interactions \cite{JM2000,DJSZ2004,CzR2006,CzR0000,MRR1990} is a conceptually simple phenomenological model for studying correlations and for a~description of magnetism and other types of electron orderings in narrow band systems with easy-plane or easy-axis magnetic anisotropy.

In this report we will focus on the zero-bandwidth limit of the extended Hubbard model with magnetic interactions for the case of arbitrary electron density \mbox{$0<n<2$}.
We consider the $U$-$J^z$ Hamiltonian of the following form:
\begin{equation}\label{row:1}
\hat{H} = U\sum_i{\hat{n}_{i\uparrow}\hat{n}_{i\downarrow}} - 2J^z\sum_{\langle i,j\rangle}{\hat{s}^z_{i}\hat{s}^z_{j}} - \mu\sum_{i}{\hat{n}_{i}},
\end{equation}
where $U$~is the on-site density interaction,  $J^{z}$~is $z$-component of the intersite magnetic exchange interaction, \mbox{$\sum_{\left\langle i,j\right\rangle}$} restricts the summation to nearest neighbours. $\hat{c}^{+}_{i\sigma}$ denotes the creation operator of an electron with spin $\sigma$ at the site $i$, \mbox{$\hat{n}_{i}=\sum_{\sigma}{\hat{n}_{i\sigma}}$}, \mbox{$\hat{n}_{i\sigma}=\hat{c}^{+}_{i\sigma}\hat{c}_{i\sigma}$} and \mbox{$\hat{s}^z_i = \frac{1}{2}(\hat{n}_{i\uparrow}-\hat{n}_{i\downarrow})$}. The chemical potential $\mu$ depending on the concentration of electrons is calculated from
\begin{equation}\label{row:2}
n = \frac{1}{N}\sum_{i}{\left\langle \hat{n}_{i} \right\rangle},
\end{equation}
with \mbox{$0\leq n \leq2$} and $N$ is the total number of lattice sites.

The model (\ref{row:1}) can be treated as an effective model of magnetically ordered insulators.
The interactions $U$ and $J^z$ will be assumed to include all the possible contributions and renormalizations like those coming from the strong electron-phonon coupling or from the coupling between electrons and other electronic subsystems in solid or chemical complexes. In such a general case arbitrary values and signs of $U$ are important to consider. We restrict ourselves to the case of positive \mbox{$J^z>0$}, because of the symmetry between ferromagnetic (\mbox{$J^z>0$}) and antiferromagnetic (\mbox{$J^z<0$}) case for lattice consisting of two interpenetrating sublattices such as for example sc or bcc lattices.

We have performed extensive study of the phase diagram of the model (\ref{row:1}) for arbitrary $n$ and $\mu$~\cite{WK2009,KKR0000}.
In the analysis we have adopted a variational approach (VA) which treats the on-site interaction $U$ exactly and the intersite interaction $J^z$ within the mean-field approximation (MFA).
We restrict ourselves to the case of the positive $J^z$, as it was mentioned above.

Let us point out that in the MFA, which does not take into account collective excitations, one obtains the same results for the \mbox{$U$-$J^z$} model  and the  \mbox{$U$-$J^{xy}$} model, where the term \mbox{$2J^z\sum{\hat{s}^z_{i}\hat{s}^z_{j}}$} is replaced with \mbox{$J^{xy}\sum{(\hat{s}^{+}_{i}\hat{s}^-_{j}+\hat{s}^{+}_{j}\hat{s}^-_{i})}$}, describing interactions between $xy$-components of spins at neighbouring sites, \mbox{$\hat{s}^{+}_i = \hat{c}^{+}_{i\uparrow}\hat{c}_{i\downarrow} = (\hat{s}^-_i)^+$}. In both cases the self-consistent equations have the same form, only the replacement \mbox{$J^{z}\rightarrow J^{xy}$} is needed and a magnetization along the $z$-axis becomes a magnetization in the $xy$-plane \cite{WK2009}.

For the model (\ref{row:1}) only the ground state phase diagram as a function of $\mu$~\cite{BS1986} and special cases of half-filling (\mbox{$n=1$})~\cite{R1979} and \mbox{$U\rightarrow\infty$}~\cite{HB1991} have been investigated till now.

Within the VA the intersite interactions are decoupled within the MFA,
what let us find a free energy per site $f(n)$.
The condition (\ref{row:2}) for the electron concentration and a~minimization of $f(n)$ with
respect to the magnetic-order parameter lead to a set of two self-consistent equations (for homogeneous phases), which are solved numerically.
The order parameter is defined as \mbox{$m^\alpha=(1/2)(m^\alpha_A+m^\alpha_B)$}, where \mbox{$m^\alpha_{\gamma}=\frac{2}{N}\sum_{i\in\gamma}{\left\langle \hat{s}_i^\alpha \right\rangle}$} is the average magnetization in a sublattice \mbox{$\gamma=A,B$} in the \mbox{$\alpha=z,xy$} direction ($s^{xy}_i$ corresponds $s^{+}_i$ here). If $m^\alpha$ is non-zero the ferromagnetic phase (F$_\alpha$) is a solution, otherwise the non-ordered phase (NO) occurs.

Phase separation (PS) is a state in which two domains with different electron concentration exist in the system
(coexistence of two homogeneous phases). The free energies of the PS states are calculated from the expression:
\begin{equation}
f_{PS}(n_{+},n_{-}) = m f_{+}(n_{+}) + (1-m) f_{-}(n_{-}),
\end{equation}
where $f_{\pm}(n_{\pm})$ are values of a free energy at $n_{\pm}$ corresponding to the
lowest energy homogeneous solutions and
\mbox{$m  = \frac{n - n_-}{n_+ - n_-}$}
is a fraction of the system with a charge density $n_+$.
We find numerically the minimum of $f_{PS}$ with respect to $n_+$ and $n_-$.

In the model considered only PS$_\alpha$ state (i.~e. a~coexistence of F$_\alpha$ and NO phases) can occur.

In the paper we have used the following convention. A~second (first) order transition is a~transition between homogeneous phases  with a~(dis-)continuous change of the order parameter at the transition temperature. A~transition between homogeneous phase and PS state is symbolically named as a~``third order'' transition. During this transition a~size of one domain in the PS state decreases continuously to zero at the~transition temperature.

Second order transitions are denoted by solid lines on phase diagrams, dotted curves denote first order transitions and dashed lines correspond to the ``third order'' transitions. We also introduce the following denotation: \mbox{$J^\alpha_0=z_1J^\alpha$} for \mbox{$\alpha=z,xy$}, where $z_1$ is the number of nearest neighbours.

Obtained phase diagrams are symmetric with respect to half-filling because of the particle-hole symmetry of the Hamiltonian (\ref{row:1}), so the diagrams will be presented only in the range \mbox{$0\leq n\leq 1$}.

\section{Results and discussion}

\subsection{The ground state}

\begin{figure}
            \centering
            \includegraphics[width=0.45\textwidth]{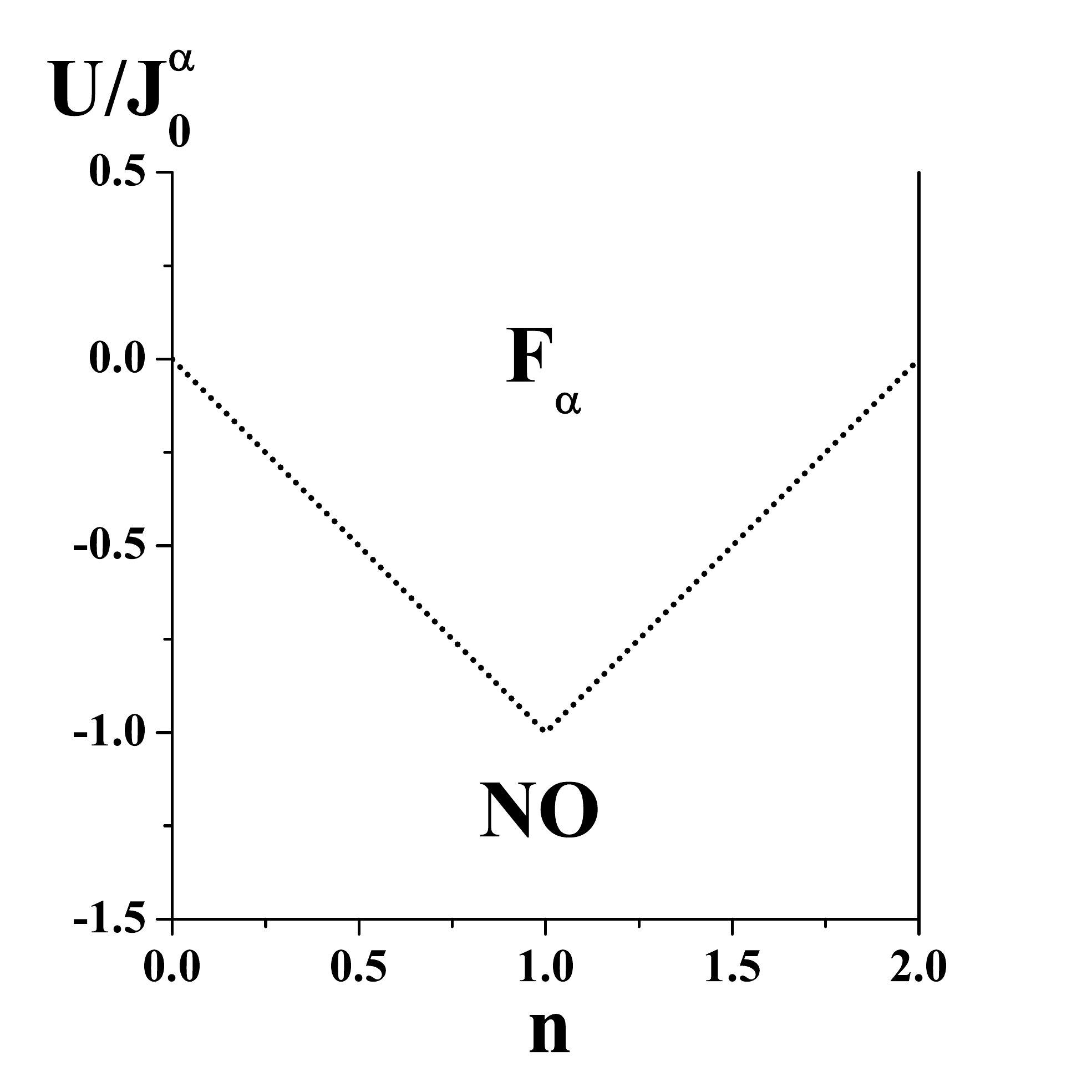}
        \caption{Ground state phase diagrams as a function of $n$ without consideration of PS states. The dotted line denotes discontinuous transition.}
        \label{rys:GDPD}
    \end{figure}

In the ground state the energies of homogeneous phases have the form: for NO: \mbox{$E_{NO}=(1/2)Un$} and for F$_{\alpha}$: \mbox{$E_F=-(1/2)J^{\alpha}_0n^2$} if \mbox{$n\leq1$} and \mbox{$E_F=U(n-1) -(1/2)J^{\alpha}_0(2-n)^2$}  if \mbox{$n\geq1$}. Comparing the energies we obtain diagram shown in Fig.~\ref{rys:GDPD}. At \mbox{$U=-J^\alpha_0(1-|n-1|)$} the first order transition F$_\alpha$--NO takes place in the system. This transition is associated with a discontinuous disappearance of the magnetization.

The first derivative of the chemical potential \mbox{$\partial \mu/\partial n= \partial^2 E/\partial n^2$} for \mbox{$U/J^\alpha_0>-1$} in the lowest energy phases is negative what implies that homogeneous phases are not stable (except \mbox{$n=1$}).

\subsection{Finite temperatures}

\begin{figure*}
            \includegraphics[width=0.45\textwidth]{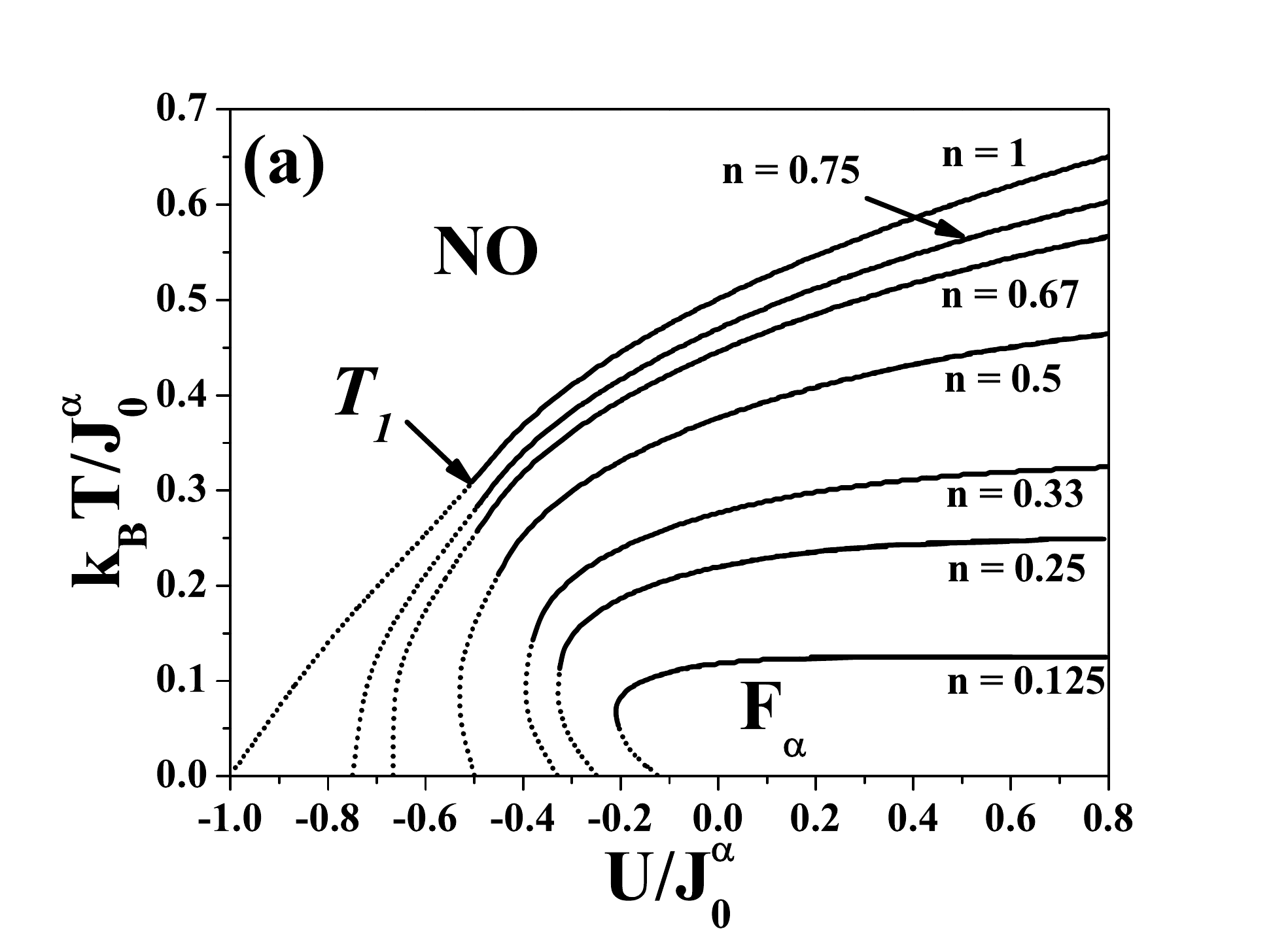}
            \includegraphics[width=0.45\textwidth]{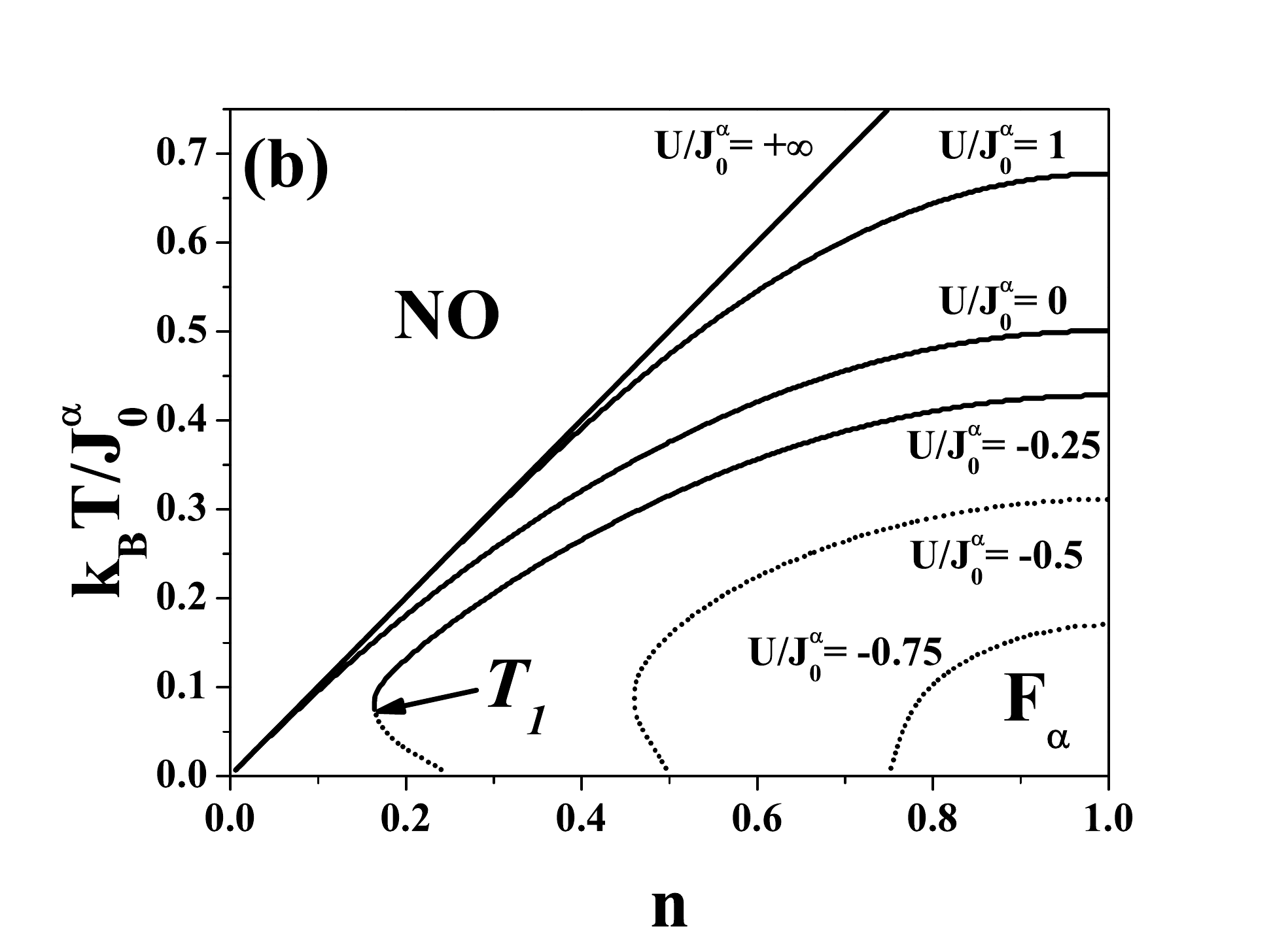}
        \caption{Phase diagrams (a) $k_BT/J^\alpha$~vs.~$U/J^\alpha_0$ for fixed $n$ and (b) $k_BT/J^\alpha_0$~vs.~$n$ for fixed $U/J^\alpha_0$  without the consideration of PS states. Dotted and solid lines denote first and second order transitions, respectively.}
        \label{rys:PDjed}
%\end{figure*}
%\begin{figure*}
            \includegraphics[width=0.45\textwidth]{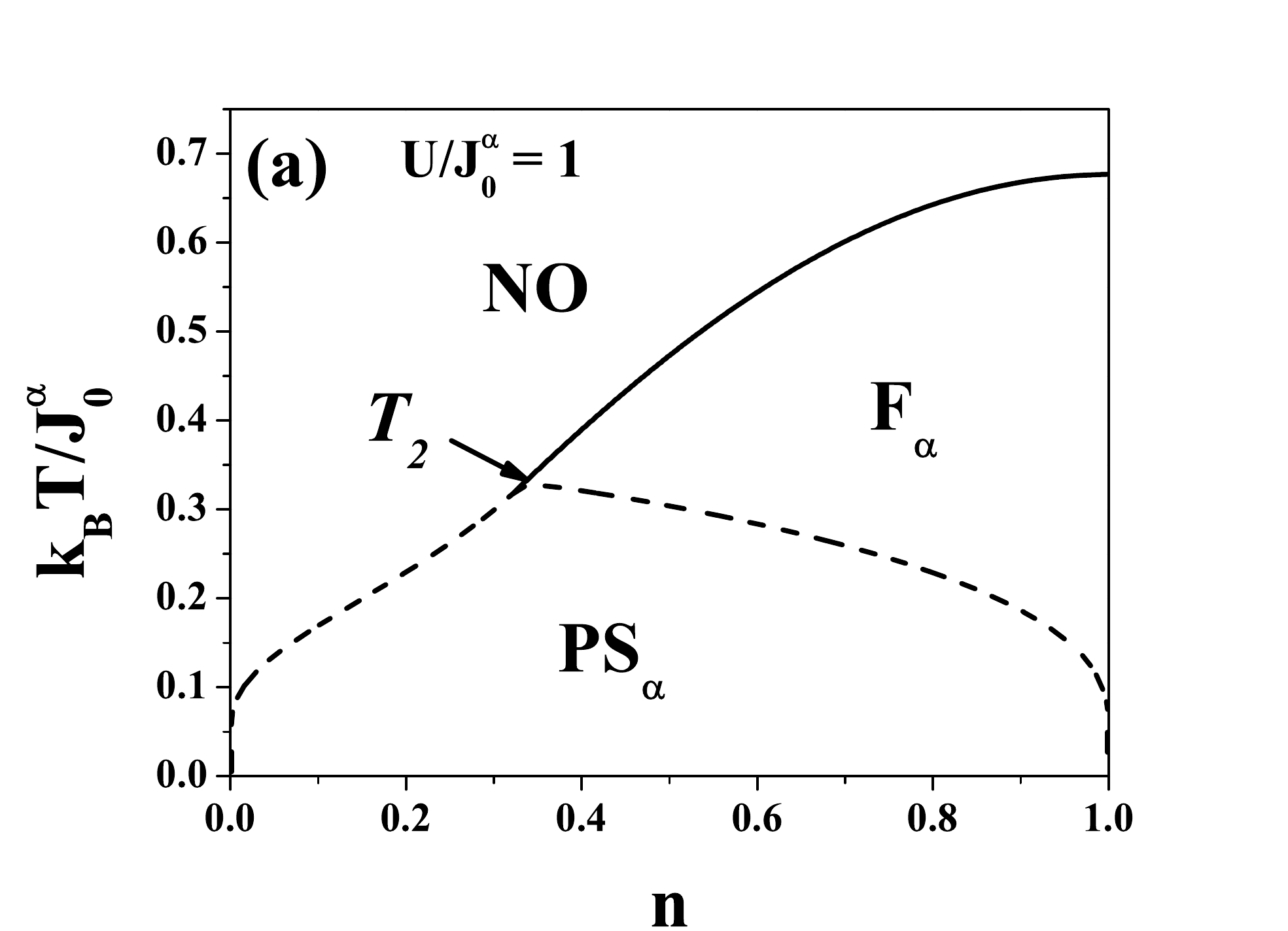}
            \includegraphics[width=0.45\textwidth]{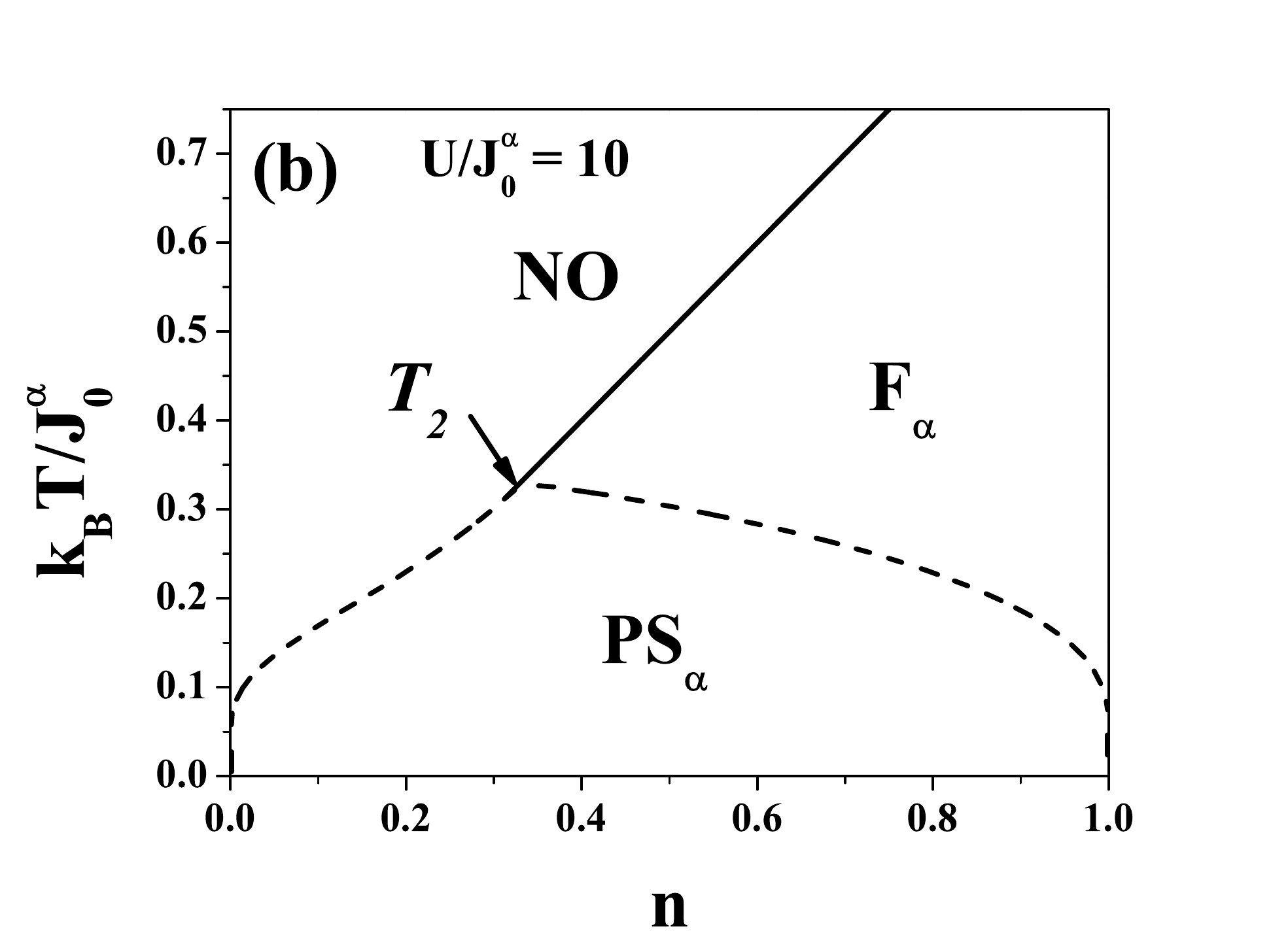}
        \caption{Phase diagrams $k_BT/J^\alpha_0$~vs.~$n$  with the consideration of PS states for: \mbox{$U/J^\alpha_0=1$} (a) and \mbox{$U/J^\alpha_0=10$} (b). Solid and dashed lines indicate second order and ``third order'' boundaries, respectively.}
        \label{rys:PDsep}
\end{figure*}

Finite temperature phase diagrams taking into account only homogeneous phases and plotted as a function of $U/J^\alpha_0$ for chosen $n$ are shown in Fig.~\ref{rys:PDjed}a.
The tricritical point $T_1$, which is connected with a~change of transition order, for \mbox{$n=1$} is located at \mbox{$k_BT/J^\alpha_0=1/3$} and \mbox{$U/J^\alpha_0=-2/3\ln2$} \cite{R1979}.

The range of  the occurrence of F$_\alpha$ phase is reduced with decreasing $n$. For \mbox{$n>0.67$} and any \mbox{$U/J^\alpha_0>-1$} we observe only one transition F$_\alpha$--NO with increasing temperature. In the range \mbox{$0.67<n\leq 1$} the $U/J^\alpha_0$ coordinate of the \mbox{$T_1$-point} remains constant, so for \mbox{$U/J^\alpha_0<-2/3\ln2$} the \mbox{F$_\alpha$--NO} transition is discontinuous. However, for \mbox{$n<0.67$} in some range of \mbox{$U/J^\alpha_0$} there can appear a sequence of two transitions: \mbox{NO--F$_\alpha$--NO}.

In Fig.~\ref{rys:PDjed}b there are shown dependencies of the transition temperature \mbox{F$_\alpha$--NO} as a function of $n$ for chosen  values of \mbox{$U/J^\alpha_0$}. The range of F$_\alpha$ stability is reduced with decreasing of \mbox{$U/J^\alpha_0$}. For \mbox{$U/J^\alpha_0>0$} and any $n$ we observe only one second order transition F$_\alpha$--NO with increasing temperature. There exist ranges of $n$ and \mbox{$U/J^\alpha_0<0$}, where the sequence of transitions: \mbox{NO--F$_\alpha$--NO} is present.

At sufficiently low temperatures homogeneous phases are not states with the lowest free energy and there PS state can occur. On the phase diagrams, where we considered the possibility of appearance of the PS states, there is a second order line at high temperatures, separating F$_\alpha$ and NO phases.  A~``third order'' transition takes place at lower temperatures, leading to a PS into F$_\alpha$ and NO phases. The critical point for the phase separation (denoted as $T_2$, a tricritical point) lies on the second order line \mbox{F$_\alpha$--NO}. Phase diagrams for \mbox{$U/J^\alpha_0=1$} and \mbox{$U/J^\alpha_0=10$} are shown in Fig.~\ref{rys:PDsep}.

In the ranges of PS stability the homogeneous phases can be metastable (if \mbox{$\partial \mu/ \partial n>0$}) or unstable (if \mbox{$\partial \mu/ \partial n<0$}). We leave a~deeper analyses of meta- and unstable states to future publications.

\section{Final remarks}

We considered a simple model for magnetically ordered insulators. It was shown that at the sufficiently low temperatures homogeneous phases do not exist and the states with phase separation are states with the lowest free energy. On phase diagrams we also observe the tricritical points, which are associated with a change of transition order (\mbox{$T_1$-point}, Fig.~\ref{rys:PDjed}) or are located in the place where the second order line connects with ``third order'' lines (\mbox{$T_2$-point}, Fig.~\ref{rys:PDsep}).

Let us stress that the knowledge of the zero-bandwidth limit can be used as starting point for a perturbation expansion in powers of the hopping and as an important test for various approximate approaches (like dynamical MFA) analyzing the corresponding finite bandwidth models.


\begin{thebibliography}{10}

\bibitem{JM2000}
G. I. Japaridze, E. Muller--Hartmann,
Phys. Rev. B, \textbf{61}, 9019 (2000).

\bibitem{DJSZ2004}
C. Dziurzik, G. I. Japaridze, A. Schadschneider, J. Zittartz,
Eur. Phys. J. B, \textbf{37}, 453 (2004).

\bibitem{CzR2006}
W. Czart, S. Robaszkiewicz,
Phys. Stat. Sol. (b) \textbf{243}, 151 (2006); Mat. Science -- Poland, \textbf{25}, 485 (2007).

\bibitem{CzR0000}
W. Czart, S. Robaszkiewicz -- in preparation.

\bibitem{MRR1990}
R. Micnas, J. Ranninger, S. Robaszkiewicz, Rev. Mod. Phys. \textbf{62}, 113 (1990).

\bibitem{WK2009}
W. K\l{}obus, Master thesis, Adam Mickiewicz University, Pozna\'n (2009).

\bibitem{KKR0000}
W. K\l{}obus, K. Kapcia, S. Robaszkiewicz -- in preparation.

\bibitem{BS1986}
U. Brandt, J. Stolze, Z. Phys. B \textbf{62}, 433 (1986).

\bibitem{R1979}
S. Robaszkiewicz,
Acta Phys. Pol. A \textbf{55}, 453 (1979); Phys. Status Solidi (b) \textbf{70}, K51 (1975).

\bibitem{HB1991}
W. Hoston, A. N. Berker, Phys. Rev. Lett. \textbf{67}, 1027
(1991).

\end{thebibliography}
\end{document}